\documentclass[twocolumn,showpacs,preprintnumbers,amsmath,amssymb]{revtex4}
\usepackage{graphicx}
\usepackage{dcolumn}% Align table columns on decimal point
\usepackage{bm}% bold math

\begin{document}

\title{Localization of 4D gravity on pure geometrical thick branes}

\author{Nandinii Barbosa--Cendejas}

\author{Alfredo Herrera--Aguilar}

\affiliation{Instituto de F\'{\i}sica y Matem\'{a}ticas, Universidad
Michoacana de San Nicol\'as de Hidalgo. \\
Edificio C--3, Ciudad Universitaria, C.P. 58040, Morelia,
Michoac\'{a}n, M\'{e}xico.\\
E-mail: nandinii@ifm.umich.mx\hspace{1cm} E-mail:
herrera@ifm.umich.mx}


\begin{abstract}
We consider the generation of thick brane configurations in a pure
geometric Weyl integrable 5D space time which constitutes a
non--Riemannian generalization of Kaluza--Klein (KK) theory. In this
framework, we show how 4D gravity can be localized on a scalar thick
brane which does not necessarily respect reflection symmetry,
generalizing in this way several previous models based on the
Randall--Sundrum (RS) system and avoiding both, the restriction to
orbifold geometries and the introduction of the branes in the action
by hand. We first obtain a thick brane solution that preserves 4D
Poincar\'e invariance and breaks $Z_2$--symmetry along the extra
dimension which, indeed, can be either compact or extended, and
supplements brane solutions previously found by other authors. In
the non--compact case, this field configuration represents a thick
brane with positive energy density centered at $y=c_2$, whereas
pairs of thick branes arise in the compact case. Remarkably, the
Weylian scalar curvature is non--singular along the fifth dimension
in the non--compact case, in contraposition to the RS thin brane
system. We also recast the wave equations of the transverse
traceless modes of the linear fluctuations of the classical
background into a Schr\"odinger's equation form with a volcano
potential of finite bottom in both the compact and the extended
cases. We solve Schr\"odinger equation for the massless zero mode
$m^2=0$ and obtain a single bound wave function which represents a
stable 4D graviton. We also get a continuum gapless spectrum of KK
states with $m^2>0$ that are suppressed at $y=c_2$ and turn
asymptotically into plane waves.
\end{abstract}

\maketitle

\section{Introduction}

In the present work we shall consider the formation of thick branes
in a particular generalization of Kaluza--Klein theory in which the
Riemannian structure of space time is enlarged into a Weylian affine
manifold where vector lengths may be not preserved along parallel
transportation. More precisely, a Weyl geometry is an affine
manifold specified by $(g_{MN},\omega_M)$, where $g_{MN}$ is the
metric tensor and $\omega_M$ is ``gauge" vector involved in the
definitions of the affine connections of the manifold. The
particular type of gauge geometries in which the gauge vector is the
gradient of a scalar function is called conformally Riemann or Weyl
integrable space time, since a conformal transformation maps a
Riemann geometry into a Weyl integrable one. If laws of physics were
invariant under conformal transformations, the Weyl scalar function
would be unobservable. However, since this is not the case, this
scalar field cannot be discarded in principle by a convenient gauge
choice; moreover, it is enough to dynamically break the conformal
invariance of a given Weyl integrable theory to transform the Weyl
scalar function into an observable field. Thus, in this scheme, a
fundamental role in the generation of thick brane configurations is
ascribed to the scalar Weyl field, which is not a bulk field.

On the other hand, the fact that we could live in a higher
dimensional space time with extended extra dimensions turns out to
be completely compatible whit present time gravitational
experiments. An interesting picture arises in such scenarios: from
the point of view of an observer located at a 3--brane in which
matter is confined, gravity is essentially 4--dimensional, however,
the world can be higher dimensional with infinite extra dimensions
and gravity can propagate in all of them. Multidimensional space
times with large extra dimensions turned out to be very useful when
addressing several problems of high energy physics like the
cosmological constant, dark matter and the mass hierarchy problem
\cite{rubakov}--\cite{lr} as well as the recent non--supersymmetric
string model realization of the Standard Model at low energy with no
extra massless matter fields \cite{kokorelis}. The striking success
of these higher dimensional scenarios motivated several
generalizations in various directions including thick brane
configurations \cite{dewolfe}--\cite{varios}. These configurations
were generalized in the framework of Weyl geometries for
$Z_2$--symmetric manifolds in \cite{ariasetal}. Moreover,
localization of 4D gravity on thick branes that break reflection
symmetry was presented in \cite{bh2} for a constant
self--interacting potential of the scalar field $(U=\lambda)$.

In this paper we keep working in a manifold endowed with Weyl
structure and present the realization of such a scenario on thick
brane solutions made out of scalar matter with a self--interacting
potential endowed with an arbitrary parameter $\xi$: $U=\lambda
e^{(1+16\xi)\omega}$, enlarging the class of potentials for which 4D
gravity can be localized.

Thus, we begin by considering a 5--dimensional Weyl gravity model in
which geometrical thick branes arise naturally without the necessity
of introducing them by hand in the action of the theory. In order to
obtain solutions which describe such configurations and respect 4D
Poincar\'e invariance we implement a conformal transformation to
pass from the Weyl frame to the Riemann one, where the Weylian
affine connections become Christoffel symbols and the field
equations are simpler, solve these equations and return to the Weyl
frame to analyze the physics of the solution. In what follows we
shall refer to this method as the conformal technique. Thus, in this
way we obtain a solution that represents a localized function which
does not necessarily respect reflection $Z_2$--symmetry and allows
for both compact and non--compact manifolds in the extra dimension.
By looking at the energy density of the scalar field of these
solutions we interpret the field configuration as thick branes. The
structure of these brane configurations depends on the topology of
the extra dimension and the value of the parameter $p(\xi)$. We
investigate as well the fluctuations of the metric around the
classical background solution to understand whether 4D gravity can
be described in our setup. We show that this is the case since the
analog quantum mechanical problem with a volcano potential for the
transverse traceless sector of the fluctuations of the metric yields
a continuum and gapless spectrum of KK states with a stable zero
mode that corresponds to the 4D graviton. We finally make our
conclusions.

Let us start by considering a pure geometrical Weyl action in five
dimensions. This non--Riemannian generalization of the Kaluza--Klein
theory is given by
\begin{equation}
\label{action} S_5^W =\int_{M_5^W}\frac{d^5x\sqrt{|g|}}{16\pi
G_5}e^{\frac{3}{2}\omega}[R+3\tilde{\xi}(\nabla\omega)^2+6U(\omega)],
\end{equation}
where $M_5^W$ is a Weyl manifold specified by the pair
$(g_{MN},\omega)$, $g_{MN}$ being the metric and $\omega$ a Weyl
scalar function. The Weylian Ricci tensor reads
$$R_{MN}=\Gamma_{MN,A}^A-\Gamma_{AM,N}^A+\Gamma_{MN}^P\Gamma_{PQ}^Q-\Gamma_{MQ}^P\Gamma_{NP}^Q,$$
where
$$\Gamma_{MN}^C=\{_{MN}^{\;C}\}-\frac{1}{2}\left(
\omega_{,M}\delta_N^C+\omega_{,N}\delta_M^C-g_{MN}\omega^{,C}\right)$$
are the affine connections on $M_5^W$, $\{_{MN}^{\;C}\}$ are the
Christoffel symbols and $M,N=0,1,2,3,5$; the constant $\tilde{\xi}$
is an arbitrary coupling parameter, and $U(\omega)$ is a
self--interaction potential for the scalar field $\omega$. This
action is of pure geometrical nature since the scalar field that
couples to gravity is precisely the scalar function $\omega$ that
enters in the definition of the affine connections of the Weyl
manifold and, thus, cannot be discarded in principle from our
consideration. Apart from the self--interaction potential, the
action (\ref{action}) is invariant under Weyl rescalings
\begin{eqnarray}
\label{weylrescalings} g'_{MN}\rightarrow\Omega^{-2}g_{MN},\qquad
\omega'\rightarrow\omega+\ln\Omega^2,\nonumber\\
\tilde{\xi}'\rightarrow\tilde{\xi}/(1+\partial_\omega\ln\Omega^2)^2,
\end{eqnarray} where $\Omega^2$ is a smooth function on $M_5^W$. Thus, from these
relations it follows that the potential must undergo the
transformation $U'\rightarrow\Omega^2U$ in order to keep such an
invariance. Thus, $U(w)=\lambda e^{\omega}$, where $\lambda$ is a
constant parameter, is the form of the potential which preserves the
scale invariance of the Weyl manifold (\ref{action}). When this
invariance is broken, the Weyl scalar field transforms from a
geometrical object into a physically observable matter degree of
freedom which, in turn, generates the thick brane configurations.

Since we are looking for a solution to the theory (\ref{action})
with 4-dimensional Poincar\'e invariance we shall consider the line
element in the form
\begin{equation}
\label{line}
ds_5^2=e^{2A(y)}\eta_{mn}dx^m dx^n+dy^2,
\end{equation}
where $e^{2A(y)}$ is the warp factor depending on the extra
coordinate $y$, and $m,n=0,1,2,3$. Thus, the 5--dimensional
stress--energy tensor is given by its 4--dimensional and pure
5--dimensional components
\begin{eqnarray}
T_{mn}=\frac{3}{8\pi G_5} e^{2A}[A''+2(A')^2]\eta_{mn}, \nonumber \\
T_{55}=\frac{6(A')^2}{8\pi G_5},
\end{eqnarray}
where the comma denotes derivatives with respect to the fifth
coordinate $y$.

\section{Solutions to the system}

Since we shall use the conformal technique to find solutions to our
system, we perform the conformal transformation
$\widehat{g}_{MN}=e^{\omega}g_{MN}$, mapping the Weylian action
(\ref{action}) into the Riemannian one
\begin{equation}
\label{confaction}
S_5^R=\int_{M_5^R}\frac{d^5x\sqrt{|\widehat
g|}}{16\pi G_5}[\widehat R+3{\xi}(\widehat\nabla\omega)^2+6\widehat
U(\omega)],
\end{equation}
where $\xi=\tilde{\xi}-1$, $\ \widehat U(\omega)=e^{-\omega}
U(\omega)$ and all hatted magnitudes and operators are defined in
the Riemann frame. In this frame we have a theory which describes
5--dimensional gravity minimally coupled to a scalar field which
possesses a self--interaction potential. After this transformation,
the line element (\ref{line}) yields the Riemannian metric
\begin{equation}
\label{conflinee} \widehat{ds}_5^2=e^{2\sigma(y)}\eta_{nm}dx^n
dx^m+e^{\omega(y)}dy^2,
\end{equation}
where $2\sigma=2A+\omega$. Further, by following \cite{ariasetal} we
introduce the new variables $X=\omega'$ and $Y=2A'$ and get the
following pair of coupled field equations from the action
(\ref{confaction})
\begin{eqnarray}
\label{fielde} X'+2YX+\frac{3}{2}X^2=\frac{1}{\xi}\frac{d\widehat
U}{d\omega}e^{\omega},\nonumber\\
Y'+2Y^2+\frac{3}{2}XY=\left(-\frac{1}{\xi}\frac{d\widehat
U}{d\omega}+4\widehat U\right)e^{\omega}.
\end{eqnarray}
In general, it is not trivial to fully integrate these field
equations. Under some assumptions, it is straightforward to
construct several particular solutions to them. However, quite often
such solutions lead to expressions of the dynamical variables that
are too complicated for an analytical treatment in closed form.

As pointed out in \cite{ariasetal}, this system of equations can be
easily solved if one uses the condition $X=kY$, where $k$ is an
arbitrary constant parameter which is not allowed to adopt the value
$k=-1$ because the system would be incompatible. It turns out that
this restriction leads to a Riemannian potential of the form
$\widehat U=\lambda e^{\frac{4k\xi}{1+k}\omega}$. Thus, under
these conditions, both field equations in (\ref{fielde}) reduce to a
single differential equation
\begin{equation}
\label{finale}
Y'+\frac{4+3k}{2}Y^2=\frac{4\lambda}{1+k}e^{(1+\frac{4k\xi}{1+k})\omega}.
\end{equation}
Those authors noticed as well that by choosing the parameter
$\xi=-(1+k)/(4k)$ (while leaving the parameter $k$ arbitrary) the
exponential function of the right hand side of (\ref{finale})
disappears and the equation can be easily solved. This case
corresponds to having a constant self--interaction potential of the
form $U=\lambda$ in the Weyl frame which, breaks the
invariance under Weyl rescalings. It is interesting to note that
after solving equation (\ref{finale}) with such a simplification,
the obtained solution $$\omega(y)=bk\ln[\cosh(ay)]$$ does not allow
the value $k=-4/3$ (apart from $k=-1$) since the constants involved
in it read $$a=\sqrt{\frac{4+3k}{1+k}2\lambda} \qquad \mbox{\rm and}
\qquad b=\frac{2}{4+3k}.$$

In this paper we shall consider another truncation that leads to a
further simplification and to a simple solution of the equation
(\ref{finale}). This can be done by setting $k=-4/3$, while leaving
$\xi$ arbitrary, allowing to have a self--interaction potential
$$U=\lambda e^{(1+16\xi)\omega}$$ in the Weyl frame. In this sense our
solution supplements the solution obtained in \cite{ariasetal},
\cite{bh2}. This potential breaks the invariance under Weyl scaling
transformations for arbitrary $\xi\ne0$ and also transforms the
geometrical scalar field $\omega$ into an observable one\footnote{In
fact, the choice of the parameter $\xi=-1/16$ leads to a constant
self--interacting potential $U=\lambda$ in the original Weyl frame,
a case which was considered by \cite{ariasetal} and \cite{bh2}.}.
Thus, after imposing the condition $k=-4/3$, the second term in the
left hand side of the equation (\ref{finale}) vanishes, yielding
\begin{equation}
Y'+12\lambda e^{p\omega}=0 \qquad \mbox{\rm or}\qquad
\omega''-16\lambda e^{p\omega}=0, \label{diffeqw}
\end{equation}
where $p=1+16\xi$.

By solving the latter equation for $\omega$ and integrating the
relation $2A'=-3\omega'/4$ one gets the following solution
\begin{eqnarray}
\label{pairsolut} \omega=-\frac{2}{p}\ln\left[\frac{\sqrt{-8\lambda
p}}{c_1}\cosh\left(c_1(y-c_2)\right)\right], \nonumber \\
e^{2A}=\left[\frac{\sqrt{-8\lambda
p}}{c_1}\cosh\left(c_1(y-c_2)\right)\right]^{\frac{3}{2p}},
\end{eqnarray}
where $c_1$ and $c_2$ are arbitrary integration constants, and we
have set to one a constant that multiplies the warp factor.

By looking at the solution, we see that for $p<0$ it constitutes a
localized object which does not necessarily preserve the reflection
$Z_2$--symmetry ($y\rightarrow -y$) along the fifth dimension and
breaks it through non--trivial values of the shift parameter $c_2$.
Thus, the 5--dimensional space time is not restricted to be an
orbifold geometry, allowing for a more general type of manifolds.
The extra coordinate can be compact or extended depending on the
signs of the constants $p$ and $\lambda$, and the real or imaginary
character of the parameter $c_1$ which, indeed, characterizes the
width of the warp factor $\Delta\sim1/c_1$. Let us consider the
cases of physical interest:

\noindent A) $\lambda>0$, $p<0$, $c_1>0$. In this case the domain of
the fifth coordinate is $-\infty<y<\infty$; thus, we have a
non--compact manifold in the extra dimension. It turns out that in
this case the warp factor is concentrated near $y=c_2$ and
represents a smooth localized function of width $\Delta$ which
remarkably reproduces the metric of the RS model in the thin brane
limit, namely, when $c_1\rightarrow\infty$, $p\rightarrow-\infty$
keeping $c_1/p=\beta$ finite.

The energy density $\mu$ of the scalar matter is given by the
null--null component of the stress--energy tensor:
\begin{eqnarray}
\mu(y)=\frac{-9c_1^2}{32p\pi G_5}\left[\frac{\sqrt{-8\lambda
p}}{c_1}\cosh\left(c_1(y-c_2)\right)\right]^{\frac{3}{2p}}\nonumber\\
\times \left[1+\frac{3-2p}{2p}\tanh^{2} \left(c_1(y-c_2)\right)\right].
\end{eqnarray}
It represents a thick brane with positive energy density\footnote{We
thank the referee of Physical Review D for pointing us the correct
interpretation of our thick bane configuration. As in \cite{bh2}, we
initially thought that we had one thick brane with positive energy
density and two branes with negative energy density, however, after
taking the thin brane limit, we realize that we have a single brane
since their parts cannot be separated at all.} centered at $c_2=2$.
In Fig. 1 we display the function $\mu$ together with its thin brane
limit, it shows a positive maximum at $y=c_2$ and a minimum at each
side of the maximum, vanishing as $y$ approaches infinity.

\begin{figure}
\includegraphics[width=8cm]{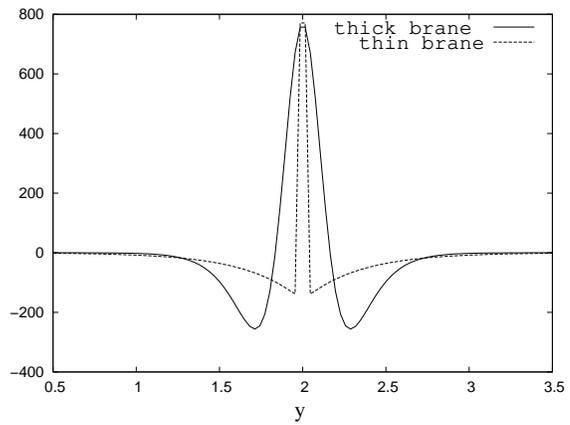}
\vspace{5mm}
\caption{\label{fig:4e3desidad}The qualitative behaviour of the
scalar energy density function $\mu$ for the non--compact case A)
and its ``reescaled" thin brane limit. The thick brane with positive
energy density is centered at $c_2=2$ and has $c_1=4$.}
\end{figure}

The 5-dimensional curvature scalar adopts the form
\begin{equation}
\label{R5nc} R_5=\frac{-6c_1^2}{p}\left[1+\frac{15-8p}{8p}
\tanh^{2}\left(c_1(y-c_2)\right)\right],
\end{equation}
and is always bounded, thus, we have a 5--dimensional manifold that
is non singular, in opposition to the RS model, where the
5--dimensional curvature scalar is singular.

\noindent B) $\lambda>0$, $p>0$, $c_1=iq_1$. In this case we get a
compact manifold along the extra dimension with $-\pi\le
q_1(y-c_2)\le\pi$. The expressions for the warp factor and the
scalar field in this compact case read
\begin{eqnarray}
e^{2A(y)}=\left[\frac{\sqrt{8\lambda
p}}{q_1}\cos\left(q_1(y-c_2)\right)\right]^{\frac{3}{2p}},\nonumber\\
\omega=-\frac{2}{p}\ln\left[\frac{\sqrt{8\lambda
p}}{q_1}\cos\left(q_1(y-c_2)\right)\right].
\end{eqnarray}
The corresponding energy density of the scalar matter is given by
\begin{eqnarray}
\mu(y)=\frac{9q_1^2}{32p\pi G_5}\left[\frac{\sqrt{8\lambda
p}}{q_1}\cos\left(q_1(y-c_2)\right)\right]^{\frac{3}{2p}}\nonumber\\
\times\left[1+\frac{2p-3}{2p}\tan^{2}\left(q_1(y-c_2)\right)\right].
\end{eqnarray}

\begin{figure}
\includegraphics[width=8cm]{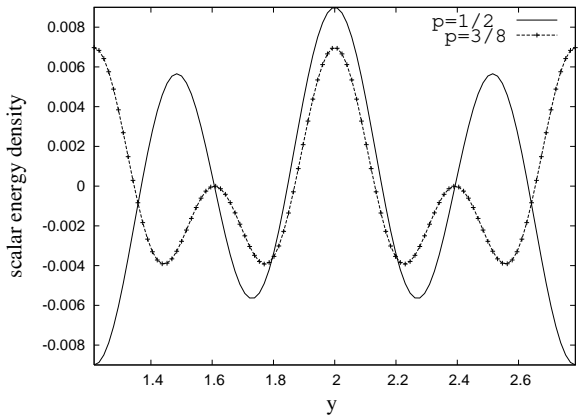}
\vspace{5mm}
\caption{\label{fig:4e3desidadcomp}The shape of the energy density
function $\mu$ for the compact case B) when $p=1/2$ and $p=3/8$ with
$c_2=2$ and $c_1=4$. Thick branes are separated at
$y=\pm\frac{\pi}{2q_1}+c_2$ due to the singular character of the
Weyl manifold at this points.}
\end{figure}

The shape of the scalar energy density $\mu$ is plotted in Fig. 2
for $p=1/2$ and $p=3/8$. The structure of the thick branes depends
on the value of the parameter $p(\xi)$. Thus, for instance, when
$3/(2p)=2n$, ($n=2,3,4,5, ...$), the shape of $\mu$ corresponds to
the case $p=3/8$, a pair of smooth thick branes of the form
displayed in Fig. 1; whereas for $3/(2p)=2n-1$, ($n=2,4,5,6, ...$),
the structure of $\mu$ is similar to the case $p=1/2$ and here we
have one thick brane of the form plotted in Fig. 1, together with
another "inverted brane", a fact that reflects the change of sign in
the warp factor when $y=\pm\frac{\pi}{2q_1}+c_2$. Actually, these
branes live in different disconnected regions of the Weyl manifold
due to the fact that the 5d curvature scalar is singular at the
points $y=\pm\frac{\pi}{2q_1}+c_2$:
\begin{equation}
\label{R5c}
R_5=\frac{6q_1^2}{p}\left[1+\frac{8p-15}{8p}\tan^{2}\left(q_1(y-c_2)\right)\right],
\end{equation}
where plausibly we have null scalar energy densities.

Other cases of physical interest are contained in A) and B), namely,
the discrete cases $\lambda>0$, $p<0$, $c_1<0$ with $p=-3/(4n)$ and
$\lambda<0$, $p<0$, $c_1<0$ (or $c_1>0$) with $p=-3/(8n)$, where
$n\ne0, n\in N$, are contained in the non--compact case A), whereas
the cases $\lambda>0$, $p>0$, $c_1=-iq_1$ with $p=3/(4n)$ and
$\lambda<0$, $p>0$, $c_1=\pm iq_1$ with $p=3/(8n)$ are included in
the case B). The remaining possible values of these parameters lead
to unphysical situations in which the warp factor and the scalar
energy density are singular at certain values of the fifth dimension
$y$ and, hence, do not represent localized functions.

\section{Fluctuations of the metric}

Let us turn to study the metric fluctuations $h_{mn}$ of the metric
(\ref{line}) given by the perturbed line element
\begin{equation}
\label{mfluct} ds_5^2=e^{2 A(y)}[\eta_{mn}+h_{mn}(x,y)]dx^m
dx^n+dy^2.
\end{equation}
Even if one cannot avoid considering fluctuations of the scalar
field when treating fluctuations of the background metric, in
\cite{dewolfe} it was shown that the transverse traceless modes of
the metric fluctuations decouples from the scalar sector and hence,
can be approached analytically.

By following this method, we perform the coordinate transformation
$dw=e^{-A}dy$, which leads to a conformally flat metric and to the
following wave equation for the transverse traceless modes
$h_{mn}^T$ of the metric fluctuations
\begin{equation}
\label{eqttm} (\partial_w^2+3A'\partial_w+\Box^{\eta})h_{mn}^T=0.
\end{equation}
This equation supports a massless and normalizable 4D graviton given
by $h_{mn}^T=C_{mn}e^{imx}$, where $C_{mn}$ are constant parameters
and $m^2=0$.

In \cite{rs} it was proved useful to recast equation (\ref{eqttm})
into Schr\"{o}dinger's equation form. In order to accomplish this,
we adopt the following ansatz for the transverse traceless modes of
the fluctuations $h_{mn}^T=e^{imx}e^{-3A/2}\Psi_{mn}(w)$ and get
\begin{equation}
\label{schrodinger} [\partial_w^2-V(w)+m^2]\Psi=0,
\end{equation}
where we have dropped the subscripts in $\Psi$, $m$ is the mass of
the KK excitation, and the potential reads
\begin{equation}
V(w)=\frac{3}{2}\partial_w^2A+\frac{9}{4}(\partial_wA)^2.
\end{equation}

For the particular non--compact case A) we have found two particular
cases ($p=-1/4$ and $p=-3/4$) for which we can invert the coordinate
transformation $dw=e^{-A}dy$ and explicitly express $y$ in terms of
$w$; although the first case is more involved, it is qualitatively
equivalent to the second one in the sense that their potential
presents a similar behaviour. For the sake of simplicity we shall
consider just the simplest case $p=-3/4$. Thus, the function $A(w)$
adopts the form
$$A(w)=\ln\left[c_1/\sqrt{c_1^4(w-w_0)^2+6\lambda}\right]$$ which
yields the following quantum mechanical potential
\begin{equation}
\label{potentialncA} V(w)=
\frac{3c_1^4\left[5c_1^4(w-w_0)^2-12\lambda\right]}{4\left[c_1^4(w-w_0)^2+6\lambda\right]^2}.
\end{equation}

In the Schr\"odinger equation, the spectrum of eigenvalues $m^2$
parameterizes the spectrum of graviton masses that a 4--dimensional
observer located at $w_0$ sees. It turns out that for the zero mode
$m^2=0$, this equation can be solved. The only normalizable
eigenfunction reads
$$\Psi_0=q\left[c_1^4(w-w_0)^2+6\lambda\right]^{-3/4},$$ where $q$ is
a normalization constant. This function represents the lowest energy
eigenfunction of the Schr\"odinger equation (\ref{schrodinger})
since it has no zeros. This fact allows for the existence of a 4D
graviton with no instabilities from transverse traceless modes with
$m^2<0$. In addition to this massless mode, there exists a tower of
higher KK modes with positive $m^2>0$.

It turns out that a similar situation takes place in the compact
case B). Remarkably, the coordinate transformation $dw=e^{-A}dy$ can
be inverted for $p=3/8$ yielding $$\cos(q_1(y-c_2))=\pm
q_1/\sqrt{q_1^2+9\lambda^2(w-w_0)^2},$$ i.e., decompactifying the
fifth dimension and pushing to infinity the singularities (an
inverse effect takes place when one compactifies the radial
coordinate $r$ in the Schwarzschild and Kerr solutions\footnote{We
are really grateful to Prof. U. Nucamendi for drawing our attention
to this effect and the correponding reference.}, see
\cite{ashtekar}). This mathematical fact implies that we actually
have two disconnected regions in the Weyl manifold: the region
$-\frac{\pi}{2}\le q_1(y-c_2)\le\frac{\pi}{2}$ is separated from the
region $\frac{\pi}{2}\le q_1(y-c_2)\le\frac{3\pi}{2}$ (since we can
shift the domain of the compact dimension to $-\frac{\pi}{2}\le
q_1(y-c_2)\le\frac{3\pi}{2}$) by the physical singularities located
at $y=\pm\frac{\pi}{2q_1}+c_2$ (recall that the curvature scalar is
singular at these points). Each one of these regions leads to
$$A(w)=\ln\left\{3\lambda/[q_1^2+9\lambda^2(w-w_0)^2]\right\}$$ and,
hence, to the following potential
\begin{equation}
\label{potentialncB}
V(w)=-\frac{27\lambda^2\left[q_1^2-36\lambda^2(w-w_0)^2\right]}{\left[q_1^2+9\lambda^2(w-w_0)^2\right]^2}
\end{equation}
and the wave function corresponding to the zero mode
$$\Psi_0=k\left[q_1^2+9\lambda^2(w-w_0)^2\right]^{-3/2},$$ where $k$
is a constant.

In Fig. 3 we display the shape of these potentials and their
respective zero mode wave functions. Both potentials have the
volcano form: a well of finite bottom and positive barriers at each
side that vanish asymptotically. The wave functions are lumps
localized around $w_0$. These facts imply that we have only one
gravitational bound state (the massless one) and a continuous and
gapless spectrum of massive KK states with $m^2>0$ in both cases A)
and B).

\begin{figure}
\includegraphics[width=8cm]{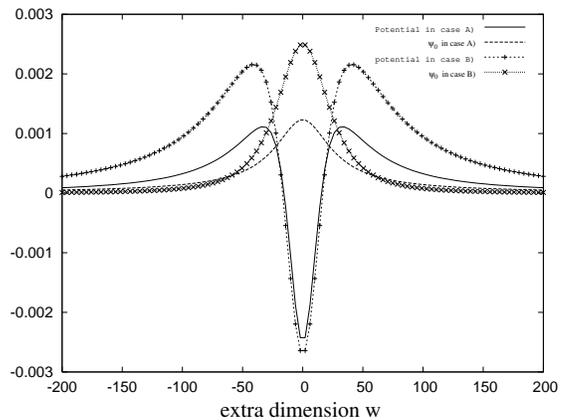}
\vspace{7mm}
\caption{\label{fig:4e3potencial}The shape of the potential $V(w)$
and the zero mode wave function $\Psi_0$ centered at $w_0=0$ for
cases A) and B).}
\end{figure}

Thus, we have obtained Weylian thick brane generalizations of
the RS model with no reflection symmetry imposed in which the 4D
effective theory possesses an energy spectrum quite similar to the
spectrum of the thin wall case, in particular, 4D gravity turns out
to be localized at a certain value of the fifth dimension in both
cases A) and B).

\section{Concluding Remarks}

We considered the formation of thick brane configurations in a
geometric Weyl integrable manifold. We used the conformal technique
to obtain a solution which preserves 4D Poincar\'e invariance and,
in particular, represents a smooth localized function characterized
by the width parameter $\Delta\sim 1/c_1$ and the constant $c_2$
which breaks the $Z_2$--symmetry along the extra dimension; both of
these parameters are integration constants of the relevant field
equation (\ref{diffeqw}), in contraposition to the solutions
obtained in \cite{ariasetal} and \cite{bh2}, where the width
parameter $\Delta\sim 1/a(\lambda,k)$ depends on the coupling
constant of the potential $U(\omega)$ and the constant $k$. Our
field configurations correspond to thick brane generalizations of
the RS model which do not restrict the 5--dimensional space time to
be an orbifold geometry, a fact that can be useful in approaching
several issues like the cosmological constant problem, black hole
physics and holographic ideas, where there is a relationship between
the position in the extra dimension and the mass scale
\cite{verlinde}. These thick branes supplement previously found
solutions with a new family in which the self--interacting potential
is endowed with an arbitrary parameter $\xi$: $U=\lambda
e^{(1+16\xi)\omega}$, enlarging the class of potentials for which 4D
gravity can be localized.

In the non--compact case A), the scalar energy density $\mu$ can be
interpreted as a generic thick brane with positive energy density
centered at $y=c_2$ and accompanied by a small amount of negative
energy density at each side; the corresponding warp factor
reproduces the metric of the RS model in the thin brane limit, even
if the matter content of the theory does not correspond to the same
brane configuration. A remarkable fact is that in this case, the
scalar curvature of the Weyl integrable manifold turns out to be
completely regular in the extra dimension. In the compact case B)
the situation is different: we have several pairs of thick brane
configurations disconnected by physical singularities. The structure
of these branes depends on the value of the parameter $p(\xi)$. In a
special case ($p=3/8$) we managed to perform a coordinate
transformation which makes the metric conformally flat,
decompactifies the fifth dimension and simultaneously pushes the
singularities of the manifold to infinity!

We wrote the wave equations of the transverse traceless modes of the
linear fluctuations of the metric into the Schr\"odinger's equation
form for both cases A) and B). The analog quantum mechanical
potential involved in it represents a volcano potential with finite
bottom: a negative well located between two finite positive barriers
that vanish when $w\longrightarrow\pm\infty$. It turned out that for
the massless zero modes ($m^2=0$) the Schr\"odinger equation can be
solved in both cases. As a result of this fact, in each case we
obtained an analytic expression for the lowest energy eigenfunction
of the Schr\"odinger equation which represents a single bound state
and allows for the existence of a stable 4D graviton since there are
no tachyonic modes with $m^2<0$. Apart from these massless states,
we also got a continuum and gapless spectrum of massive KK modes
with positive $m^2>0$ that are suppressed at $y=c_2$ and turn
asymptotically into continuum plane waves in both cases A) and B),
as in \cite{lr}, \cite{dewolfe} and \cite{bh2}.

The shape of the analog quantum mechanical potential and the
localization of 4D gravity on thick branes with a continuum and
gapless spectrum of massive KK modes are quite similar to those
obtained by \cite{lr}, \cite{dewolfe} and \cite{gremm}.

{\bf Acknowledgements.} The authors are really grateful to A.
Merzon, U. Nucamendi, C. Schubert and T. Zannias for fruitful
discussions while this investigation was carried out and acknowledge
as well useful discussions with A. Bashir, A. Corichi, V. Manko and
H. Quevedo. This research was supported by grants CIC-UMSNH-4.16 and
CONACYT-F42064.

%************************************************************************

\end{document}